# Designing a Custom Chaos Engineering Framework for Enhanced System Resilience at Softtech

Ethem Utku Aktas
utku.aktas@softtech.com.tr
Softtech Inc. R&D Center
Istanbul, Turkey

Burak Tuzlutas
burak.tuzlutas@softtech.com.tr
Softtech Inc. R&D Center
Istanbul, Turkey

Burak Yesiltas
burak.yesiltas@softtech.com.tr
Softtech Inc. R&D Center
Istanbul, Turkey

## Abstract

Chaos Engineering is a discipline which enhances software resilience by introducing faults to observe and improve system behavior intentionally. This paper presents a design proposal for a customized Chaos Engineering framework tailored for Softtech, a leading software development company serving the financial sector. It outlines foundational concepts and activities for introducing Chaos Engineering within Softtech, while considering financial sector regulations. Building on these principles, the framework aims to be iterative and scalable, enabling development teams to progressively improve their practices. The study addresses two primary questions: how Softtech's unique infrastructure, business priorities, and organizational context shape the customization of its Chaos Engineering framework and what key activities and components are necessary for creating an effective framework tailored to Softtech's needs.

## 1 Introduction

In today's rapidly evolving technological environment, ensuring the reliability and resilience of software systems is critical. As systems grow in complexity, the potential for unexpected failures increases, posing risks to business operations and customer satisfaction. Traditional testing methods often fall short in identifying and mitigating these risks, as they typically do not account for the unpredictable nature of real-world environments.

Chaos engineering has emerged as a proactive approach to address this challenge. By intentionally injecting failures into a system, chaos engineering aims to uncover weaknesses and improve system robustness before actual incidents occur. This practice is grounded in experimentation and empirical evidence, allowing organizations to gain a deeper understanding of their systems' behavior under stress [2].

The concept of chaos engineering was popularized by Netflix, which developed the Chaos Monkey tool to test the resilience of its streaming service [26]. Since then, the practice has gained traction across various industries, with companies leveraging it to enhance their systems' fault tolerance and operational stability [11, 12, 26]. The benefits include increased availability, lower mean time to resolution (MTTR), lower mean time to detection (MTTD), fewer bugs shipped to product, and fewer outages [9].

Softtech[1] is a leading software development company in Turkey, primarily serving IsBank[2], the largest private bank in the country. Softtech develops a wide range of software solutions, including banking applications and enterprise software systems, with the aim of delivering robust, scalable, and secure software. Softtech's infrastructure presents unique challenges due to the stringent regulatory compliance requirements in the financial sector, such as data privacy measures or regulatory workflows. These factors necessitate a tailored approach to chaos engineering. This paper serves as a design proposal for applying chaos engineering at Softtech, outlining the framework's foundational concepts. Although empirical validation and implementation are planned for future work, this proposal sets the stage for further exploration and validation.

The study seeks to answer two primary questions:

- RQ1: How can Softtech's unique infrastructure, business priorities, and organizational context shape the customization of its Chaos Engineering framework?
- RQ2: What are the key activities and components necessary for creating an effective Chaos Engineering framework tailored to Softtech's needs?

This paper is structured as follows: In the next section, we provide the related work on Chaos Engineering, including its principles, existing frameworks, tools, techniques, and challenges for the financial sector. Section 3 details the design foundations of the Chaos Engineering framework for Softtech, identifying specific requirements, proposed framework, and core components. Section 4 summarizes the answers to the research questions. Finally, Section 5 concludes the paper and discusses future work.

---
[1]https://softtech.com.tr/
[2]https://www.isbank.com.tr/

## 2 Related Work

### 2.1 Principles of Chaos Engineering

Chaos Engineering has emerged as a systematic approach to building confidence in distributed system capabilities by introducing controlled experiments that simulate the real world. The fundamental principles are rooted in scientific methodology and empirical observation.

The core principle is the deliberate introduction of controlled disruption to verify system behavior. This approach differs from traditional testing methodologies by focusing on proactive failure injection rather than reactive problem-solving. According to recent academic research [22], the practice builds upon the following essential principles:

- **Building a Hypothesis:** Engineers start by defining "steady state" behavior that indicates normal system operation. This baseline serves as the control group for experiments.
- **Setting up Monitoring:** Tools are used to monitor the system for deviations from its normal behavior during the controlled experiments.
- **Conducting Experiments:** Real-world systems face unpredictable conditions. Chaos experiments are designed to reflect this by varying real-world events like server failures, malformed responses, or network latency. Results are then reviewed, improvements applied and tests repeated.
- **Minimizing Impact:** To avoid widespread outages, experiments are designed to minimize their impact, often starting small and gradually increasing in scope.
- **Expanding the Experiments to Production:** While controversial, the most valuable insights come from experimenting in production environments, where actual user behavior and data patterns exist.
- **Automation and Continuous Learning:** Automating the experiments for repeatability and continuously analyzing results is crucial to enhance system resilience.

### 2.2 Existing Chaos Engineering Frameworks

Recent studies have identified multiple frameworks that approach chaos experimentation from different angles [11], each offering unique features and capabilities. By leveraging these resources, organizations can systematically identify and address potential weaknesses in their systems, enhancing reliability and performance.

The pioneer in this field, *Netflix's Simian Army*, introduced the concept of chaos agents ("monkeys") that systematically inject failures into production systems. The most famous component, *Chaos Monkey* [20], randomly terminates instances in production to ensure service resilience. *Chaos Mesh* [17] is an open-source chaos engineering platform focused on Kubernetes environments. It provides fine-grained chaos experiment management and supports multiple types of fault injection. *ChaosTwin* [23, 24] is a novel management framework that combines chaos engineering principles with digital twin technology, enabling organizations to conduct experiments in a simulated environment before moving to production. *Chess* Framework [16, 19] is a systematic evaluation framework designed specifically for self-adaptive systems, incorporating chaos engineering principles to assess system resilience and recovery capabilities.

### 2.3 Tools and Techniques in Chaos Engineering

The implementation of chaos engineering requires a sophisticated toolkit and well-defined techniques [7]. Modern practitioners employ various approaches and tools to ensure comprehensive system testing and validation. Key implementation techniques are:

- **Fault Injection** Systematic introduction of failures at various system layers to observe behavior and recovery patterns. These include:
  - *Network-Level Chaos:* Latency injection, Packet loss simulation, Bandwidth throttling, DNS failure simulation,
  - *Resource-Level Chaos:* CPU stress testing, Memory exhaustion, Disk I/O saturation, Storage corruption simulation,
  - *Application-Level Chaos:* Service dependency failure, API error injection, Database connection termination, Cache invalidation.
- **Chaos Experimentation Platforms:** Modern tools like Gremlin [10], LitmusChaos [15] and Chaos Mesh [17] provide automated frameworks for executing and monitoring chaos experiments, making the process more scalable and consistent.
- **Resilience Measurement:** Quantitative assessment of system recovery capabilities through metrics like Mean Time To Recovery (MTTR) and Service Level Objectives (SLOs).
- **Game Days:** Organized events where teams deliberately trigger failures and practice response procedures [9].

The field continues to evolve, with new tools and methodologies emerging [27] that focus on cloud-native environments and automated resilience testing.

### 2.4 Challenges in Customizing Chaos Engineering for the Financial Sector

The current literature lacks specific, actionable examples or in-depth technical and process guidance for customizing chaos engineering within large, regulated financial institutions. To the best of our knowledge, no paper directly documents real-world chaos engineering customizations for on-premise deployments, or deep compliance and data-privacy integration. While some sources mention banking organizations as use cases for fault-injection platforms [6], they do not provide the necessary technical or process details about actual financial-sector implementations. Existing works focus on cloud-native environments and general DevOps frameworks [1, 4–6, 12, 13], without addressing the specific requirements of regulated financial deployments, including compliance, data privacy, and role-based approvals. For actual finance-sector best practices in these areas, practitioners need to look beyond the current academic and practitioner publications, or seek confidential/industry-shared case materials.

## 3 Design Foundations for Chaos Engineering at Softtech

### 3.1 Identifying Softtech's Specific Requirements

*3.1.1 Technical Infrastructure.* Softtech offers tailored solutions for IsBank and other clients, leveraging expertise in COBOL, Java, C#, and C++. It supports platforms from mainframe systems to mobile technology, emphasizing security and data privacy. Softtech developed Plateau [3], an open-source, low-code microservice framework on Kubernetes, to meet IsBank's needs and serve other customers. This platform includes components like Plateau Process Manager for process management and ProEmtia for industrial trade, which could be initial subjects for chaos experiments.

*3.1.2 Business Priorities.* Softtech's *commitment to zero-downtime* for core banking services is crucial for maintaining customer trust and operational efficiency. Chaos engineering helps achieve this by simulating failures like server crashes and network outages, identifying and mitigating potential issues to ensure continuous availability. *Regulatory compliance* in the financial sector is also vital, and chaos engineering tests the effectiveness of security controls by stressing data encryption, access controls, and audit logging. Additionally, *time-sensitive transaction processing* is critical to prevent financial losses and customer dissatisfaction.

*3.1.3 Organizational Context.* Softtech's *DevOps-oriented structure* promotes collaboration between development and operations, enhancing adaptability and quick delivery through Agile and lean methods that support CI/CD. A robust *monitoring infrastructure*, with tools like Prometheus and Grafana, ensures system health, while *incident response procedures* minimize disruptions. The *diverse team*, skilled in various technologies, benefits from continuous training to tackle challenges and deliver innovative solutions.

### 3.2 Activities in the Proposed Framework

Based on specific requirements of Softtech, we focus on the four key activities, as proposed by Jernberg [11]: *Discovery, Implementation, Sophistication, and Expansion.* Furthermore, we plan to focus on the Plateau platform at Softtech, as it is Kubernetes-based and central to Softtech's future architecture.

In the *Discovery* phase, we'll define a backlog of chaos experiments tailored for the Plateau platform by analyzing the platform's architecture and dependencies. Collaboration with development, operations, and security teams will be necessary to help identify critical components and prioritize areas essential for Softtech's zero-downtime requirements.

The *Implementation* phase involves executing a selected chaos experiment from the backlog using tools like LitmusChaos. This experiment will target a selected component or service within the Plateau platform.

The *Sophistication* step refines and automates Chaos Engineering experiments for production environments, using the Chaos Maturity Model [3] to enhance resilience testing. It ensures steady-state definitions cover both technical and business metrics. Regular sophistication activities help teams incrementally improve their Chaos Engineering practices, resulting in more resilient systems.

The *Expansion* phase adds more chaos experiments to the current practice, broadening the resilience testing scope on the Plateau platform. New experiments will target additional components, addressing a wider range of failure scenarios to enhance platform resilience. Engaging various teams within Softtech fosters a collaborative culture and promotes the sharing of findings and best practices, ensuring continuous improvement.

*3.2.1 Regulatory Compliance.* Our Chaos Engineering framework integrates compliance validation processes to meet stringent financial sector regulations. Key measures include:

- *Non-intrusive controls* ensure experiments have minimal impact, are reversible, safe, and transparent, maintaining system integrity within the framework.
- *Logging and monitoring* involve robust action tracking and secure log storage, with integration into monitoring tools for continuous compliance oversight.
- *Audits and compliance* involve regular checks with automated scripts to validate data encryption and access controls, ensuring full compliance throughout experiments.
- *Compliance Reporting* provides detailed reports for audits, highlighting compliance checks, issues, and resolutions, supporting regulatory adherence.

Stakeholder involvement, particularly from compliance officers and security teams, is vital for aligning the framework with regulatory demands. Future plans include exploring advanced technologies to enhance compliance monitoring and reporting, ensuring our practices improve resilience while maintaining system integrity and security.

### 3.3 Core Components

To develop a Chaos Engineering framework for Softtech, we chose LitmusChaos as the primary chaos orchestration tool. Its Kubernetes-native capabilities for planning and executing chaos experiments make it ideal for the Plateau platform. Despite alternatives like Gremlin and Chaos Mesh, LitmusChaos was selected for its seamless Kubernetes integration, community support, and extensibility, aligning with our infrastructure and scalability needs.

Within LitmusChaos, a comprehensive set of predefined chaos scenarios will be created to address failure modes like network latency, pod failures, and resource exhaustion. Future validation will track key metrics—such as MTTR, service availability, and error rates—to evaluate the impact on Softtech's SLAs and system performance, guiding iterative framework improvements.

The framework will use advanced monitoring tools to evaluate chaos experiments. *Prometheus and Grafana* [8, 25] will provide real-time system insights, while *Kiali* [14] will visualize service mesh interactions and performance. *New Relic* [21] will offer application performance monitoring for in-depth analysis during chaos tests.

LitmusChaos will include automated rollback mechanisms for stability when systems don't recover as expected. A reporting module will provide detailed experiment outcomes, helping assess resilience and identify improvement areas. MongoDB [18] will store chaos experiment data, enabling efficient querying and analysis of historical information.

---

[3]https://softtech.com.tr/en/plateau-platform-en/

## 4 Summary of Research Questions and Answers

> **Research Question 1**
>
> **How can Softtech's unique infrastructure, business priorities, and organizational context shape the customization of its Chaos Engineering framework?**
> Softtech's unique infrastructure and its diverse platform support, mandates a Chaos Engineering framework that accommodates various technologies. The use of its open-source, cloud-based Kubernetes platform, Plateau, as the foundation integrates the system's current and future needs. Tailoring this framework requires simulating failures that aligns with zero-downtime goals, regulatory compliance, and the company's DevOps structure. It integrates with CI/CD pipelines and leverages monitoring tools.

> **Research Question 2**
>
> **What are the key activities and components necessary for creating an effective Chaos Engineering framework tailored to Softtech's needs?**
> Key activities include Discovery, Implementation, Sophistication, and Expansion. Discovery identifies the application under test, relevant experiments and prioritizes them. Implementation uses tools like LitmusChaos integrated with CI/CD pipelines. Sophistication refines and automates experiments, while Expansion adds more experiments and updates tool configuration. Core components include predefined scenarios, monitoring tools (Prometheus, Grafana, Kiali), New Relic for performance monitoring, automated rollbacks, and a reporting module with MongoDB.

## 5 Conclusion and Future Work

In conclusion, the customized Chaos Engineering framework designed for Softtech leverages its unique technical infrastructure, business priorities, and organizational context. By integrating tools within CI/CD pipelines, the framework aims to enhance system resilience through progressive and iterative improvements. This tailored approach supports Softtech's goals of zero-downtime and regulatory compliance, ensuring a robust and scalable solution for improved resilience testing and incident reduction.

While this paper primarily focuses on the design proposal, we acknowledge the importance of empirical validation. Future work will focus on the practical implementation of this framework across various applications/products within Softtech. This includes conducting pilot tests to validate its effectiveness, gathering qualitative feedback from stakeholders to ensure the design meets their needs, and training teams on the effective use of the framework. Gradually, we will scale the implementation across the entire organization. Continuous feedback will be gathered to refine and optimize the framework, ensuring it meets the evolving needs of the company. By addressing these aspects, we aim to provide a comprehensive and validated approach to chaos engineering at Softtech.

## 6 Acknowledgments

## References

[1] Merishani Arsecularatne and Ruwan Wickramarachchi. 2023. Adoptability of Chaos Engineering with DevOps to Stimulate the Software Delivery Performance. In *2023 International Research Conference on Smart Computing and Systems Engineering (SCSE)*, Vol. 6. IEEE, 1–8.
[2] Ali Basiri, Niosha Behnam, Ruud De Rooij, Lorin Hochstein, Luke Kosewski, Justin Reynolds, and Casey Rosenthal. 2016. Chaos engineering. *IEEE Software* 33, 3 (2016), 35–41.
[3] Ali Basiri, Aaron Blohowiak, Lorin Hochstein, Nora Jones, and Casey Rosenthal. 2017. *Chaos Engineering: Building Confidence in System Behavior Through Experiments*. O'Reilly Media, Inc.
[4] Ali Basiri, Lorin Hochstein, Nora Jones, and Haley Tucker. 2019. Automating chaos experiments in production. In *2019 IEEE/ACM 41st International Conference on Software Engineering: Software Engineering in Practice (ICSE-SEIP)*. IEEE, 31–40.
[5] Aaron Blohowiak, Ali Basiri, Lorin Hochstein, and Casey Rosenthal. 2016. A platform for automating chaos experiments. In *2016 IEEE International Symposium on Software Reliability Engineering Workshops (ISSREW)*. IEEE, 5–8.
[6] Carlos Camacho, Pablo C Cañizares, Luis Llana, and Alberto Núñez. 2022. Chaos as a Software Product Line—a platform for improving open hybrid-cloud systems resiliency. *Software: Practice and Experience* 52, 7 (2022), 1581–1614.
[7] Navdeep Singh Gill. 2024. Chaos engineering: tools, principles and best practices. https://www.xenonstack.com/insights/chaos-engineering. Accessed: 2025-02-23.
[8] Grafana Labs. 2023. Grafana. https://grafana.com/.
[9] Gremlin. 2023. Chaos engineering: the history, principles, and practice. https://www.gremlin.com/community/tutorials/chaos-engineering-the-history-principles-and-practice. Accessed: 2024-11-18.
[10] Gremlin Inc. 2025. Gremlin. https://www.gremlin.com/. Accessed: 2025-11-12.
[11] Hugo Jernberg. 2020. Building a Framework for Chaos Engineering. *LU-CS-EX* (2020).
[12] Hugo Jernberg, Per Runeson, and Emelie Engström. 2020. Getting Started with Chaos Engineering-design of an implementation framework in practice. In *Proceedings of the 14th ACM/IEEE International Symposium on Empirical Software Engineering and Measurement (ESEM)*. 1–10.
[13] Dominik Kesim, André van Hoorn, Sebastian Frank, and Matthias Häussler. 2020. Identifying and prioritizing chaos experiments by using established risk analysis techniques. In *2020 IEEE 31st International Symposium on Software Reliability Engineering (ISSRE)*. IEEE, 229–240.
[14] Kiali Authors. 2023. Kiali. https://kiali.io/.
[15] LitmusChaos. 2020. LitmusChaos - open source chaos engineering platform. https://litmuschaos.io/. Accessed: 2024-12-10.
[16] Sehrish Malik, Moeen Ali Naqvi, and Leon Moonen. 2023. Chess: A framework for evaluation of self-adaptive systems based on chaos engineering. In *2023 IEEE/ACM 18th Symposium on Software Engineering for Adaptive and Self-Managing Systems (SEAMS)*. IEEE, 195–201.
[17] Chaos Mesh. 2025. Chaos Mesh: a powerful chaos engineering platform for kubernetes. https://chaos-mesh.org/docs. Accessed: 2025-03-02.
[18] MongoDB, Inc. 2023. MongoDB. https://www.mongodb.com/.
[19] Moeen Ali Naqvi, Sehrish Malik, Merve Astekin, and Leon Moonen. 2022. On evaluating self-adaptive and self-healing systems using chaos engineering. In *2022 IEEE international conference on autonomic computing and self-organizing systems (ACSOS)*. IEEE, 1–10.
[20] Netflix. 2025. Chaos Monkey. https://netflix.github.io/chaosmonkey/. Accessed: 2025-01-10.
[21] New Relic, Inc. 2023. New Relic. https://newrelic.com/.
[22] Joshua Owotogbe, Indika Kumara, Willem-Jan Van Den Heuvel, and Damian Andrew Tamburri. 2024. Chaos Engineering: A Multi-Vocal Literature Review. *arXiv preprint arXiv:2412.01416* (2024).
[23] Filippo Poltronieri, Mauro Tortonesi, and Cesare Stefanelli. 2021. Chaostwin: A chaos engineering and digital twin approach for the design of resilient it services. In *2021 17th International Conference on Network and Service Management (CNSM)*. IEEE, 234–238.
[24] Filippo Poltronieri, Mauro Tortonesi, and Cesare Stefanelli. 2022. A chaos engineering approach for improving the resiliency of it services configurations. In *NOMS 2022-2022 IEEE/IFIP Network Operations and Management Symposium*. IEEE, 1–6.
[25] Prometheus Authors. 2023. Prometheus. https://prometheus.io/.
[26] Casey Rosenthal and Nora Jones. 2020. *Chaos engineering: system resiliency in practice*. O'Reilly Media.
[27] Rahul Singh. 2024. Chaos Engineering tools in 2024. https://www.devopsschool.com/blog/chaos-engineering-tools-in-2024/. Accessed: 2024-12-21.